\documentclass[twocolumn,floatfix,prx,aps]{revtex4-1}
\usepackage[T1]{fontenc}
\usepackage[latin9]{inputenc}
\usepackage{bm}
\usepackage{amsmath}
\usepackage{amssymb}
\usepackage{graphicx}
\usepackage{esint}
\usepackage[unicode=true, bookmarks=false, breaklinks=false,pdfborder={0 0 1},backref=false,colorlinks=false] {hyperref}

\makeatletter

\usepackage{color}
\usepackage{bm}
\usepackage{bbold}
\usepackage{soul}
\usepackage{epsfig}
\usepackage{verbatim}
\usepackage{array}

\newcommand{\be}{\begin{equation}}
\newcommand{\ee}{\end{equation}}
\newcommand{\bea}{\begin{eqnarray}}
\newcommand{\eea}{\end{eqnarray}}

\makeatother

\begin{document}
\title{Multi-particle interferometry in the time-energy domain with localized topological quasiparticles}
\author{Alex Zazunov,$^{1}$ Reinhold Egger,$^{1}$ and Yuval Gefen$^{2}$\ }
\affiliation{
 $^{1}$~Institut f\"ur Theoretische Physik, Heinrich-Heine-Universit\"at,
D-40225 D\"usseldorf, Germany\\
$^{2}$~Department of Condensed Matter Physics, Weizmann Institute,
7610001 Rehovot, Israel}

\date{\today}

\begin{abstract}
We propose  multi-particle interference protocols in the time-energy domain which are able to probe
localized topological quasiparticles.
Using a set of quantum dots tunnel-coupled to a topologically nontrivial system,
the time dependence of the dot level energies defines a many-body interferometry platform 
which (to some extent) is similar to the Hong-Ou-Mandel (HOM) interferometer.
We demonstrate that for a superconducting island harboring at least four Majorana bound states, 
the probability distribution of the final dot occupation numbers 
will exhibit a characteristic interferometric pattern with robust and quantized $\pi$ phase shifts. 
This pattern is shown to be qualitatively different for topologically trivial variants of our setup. 
Apart from identifying the presence of topological quasiparticles, the interferometer can be used to manipulate the 
quantum state in the topologically nontrivial sector by means of post-selection.  
\end{abstract}

\maketitle
\section{Introduction}

Interferometry is a key concept of ubiquitous appearance in physics \cite{Hariharan2007}. Using different types of interferometers, numerous otherwise inaccessible insights have been obtained in atomic physics, quantum optics, astronomy, general relativity, and  
condensed matter physics. In particular, interferometry provides information about quantum coherence, quantum correlations, and the exchange statistics of many-particle systems.  
For instance, in the celebrated Hong-Ou-Mandel (HOM) interferometer \cite{Hong1987}, two particles are emitted from phase-uncorrelated inputs and impinge 
on a $1/2$ beam splitter. The arrival coincidence measured at two separate outputs then 
probes the indistinguishability and the quantum statistics of the outgoing particles \cite{Hong1987,HBT,Neder2007,Bocquillon2013,Freulon2015,Bauerle2018}. Theoretical work has addressed both normal \cite{Samuelsson2004,Moskalets2008,Jonckheere2012} and superconducting \cite{Beenakker2014,Ferraro2015} systems. 
HOM interferometry has been demonstrated long ago for photons \cite{Hong1987}, and more recently also for electrons in solid-state devices \cite{Neder2007,Bocquillon2013,Freulon2015}.  Unfortunately, interferometry involving topological quasiparticles \cite{Nayak2008,Hasan2010,Wen2017}, e.g., anyons in the fractional quantum Hall regime \cite{Campagnano2012,Rosenow2016,Barbarino2019} or chiral Majorana edge modes in a topological superconductor (TS) \cite{Fu2009,Akhmerov2009}, has so far
remained challenging (but see Ref.~\cite{Manfra2019}).   Moreover, for \emph{spatially localized} topological quasiparticles such as 
Majorana bound states (MBSs) \cite{Alicea2012,Lutchyn2017,Ironbased1,Ironbased2,Ironbased3}, traditional interferometric approaches are not directly applicable. 

We here target a different class of platforms for realizing multi-particle interferometry: interference protocols in the time-energy domain. 
Our protocols are able to probe
localized topological quasiparticles in systems tunnel-coupled to a set of $N$ electronic terminals.
We demonstrate the feasibility of such an approach by analyzing the interference dynamics in a system with multiple MBSs, where the terminals are represented by
single-level quantum dots with time-dependent occupation numbers, ${\bm n}(t)=(n_1\cdots n_N)$ with $n_j=0,1$. 
One then runs a time-dependent protocol for the dot energy levels, $\varepsilon_j(t)$, such that electrons can enter or leave the system through a stochastic sequence of non-adiabatic 
transitions of Landau-Zener (LZ) type  \cite{LZ1,LZ2,LZ3,LZ4,Shimshoni1991,LZreview}.   
As made precise below, this sequence implements interfering trajectories in the time-energy domain where the interfering entities are composite particles obtained by fusing electrons and topological quasiparticles. For the case of MBSs, the Majorana operator algebra results in effective spin-$1/2$ particles.  Starting at time $t_i$ from an initial state with dot occupation numbers $\bm n^i={\bm n}(t_i)$, one measures all electron occupation numbers  upon completion of the protocol,  ${\bm n}^f={\bm n}(t_f)$. By repeating this protocol many times for the same initial dot configuration $\bm n^i$,  one obtains the probability
distribution $P[{\bm n}^f]$  (the dependence on $\bm n^i$ is kept implicit). The latter distribution represents the key object of interest. We show below that it encodes a nontrivial interference signal for selected final dot configurations ${\bm n}^f$.

While our approach is inspired by the HOM setup, there are
several major differences. First, instead of the time-space domain, this interferometer operates in the time-energy domain defined by the protocol
$\{ \varepsilon_j(t)\}$.  Second, particle number does not have to be conserved between the emission and detection times, and nontrivial interference signatures, e.g., a robust and quantized $\pi$ phase shift, can be traced back to this feature, see Sec.~\ref{sec3b}. In fact, an effectively particle-number conserving variant of our setup, see Sec.~\ref{sec4c}, does not exhibit such $\pi$ phase shifts.
Third, instead of employing shot noise measurements as is possible for chiral edge modes,  electrons  are injected from, and measured in, quantum dots by means of standard charge sensing techniques. 
Finally,  the interfering trajectories can be understood in terms of  composite objects built from electrons and topological quasiparticles.  While our scheme probes interference properties of
localized topological quasiparticles indirectly, a key advantage is that only  electron states need to be prepared and read out.    The experimental challenges are therefore comparatively modest.

\begin{figure}
\centering
\includegraphics[width=0.9\columnwidth]{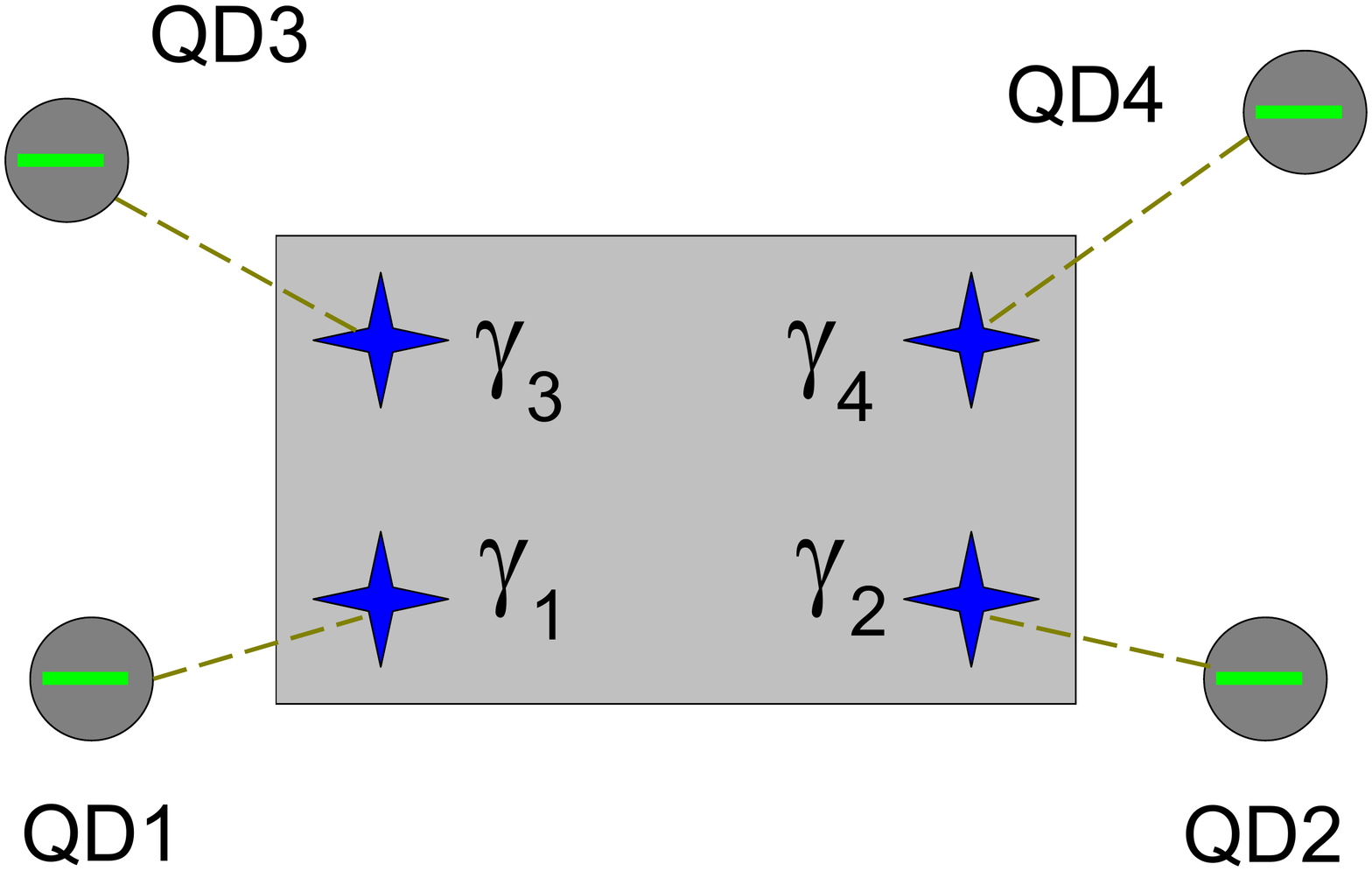}
\vspace{-1cm}
\caption{\label{fig1}  Schematic setup for a multi-particle interferometer in the time-energy domain, where 
$N=4$ single-level quantum dots are tunnel coupled (dashed lines) to a TS island (grey box) harboring MBSs described by operators $\gamma_{j}$.
The time-dependent dot level energies $\varepsilon_j(t)$ can be controlled by gate voltages.  
Readout devices for measuring the final dot occupation numbers $\bm n^f$ 
are not shown.  }
\end{figure}

\begin{figure}
\centering
\includegraphics[width=0.9\columnwidth]{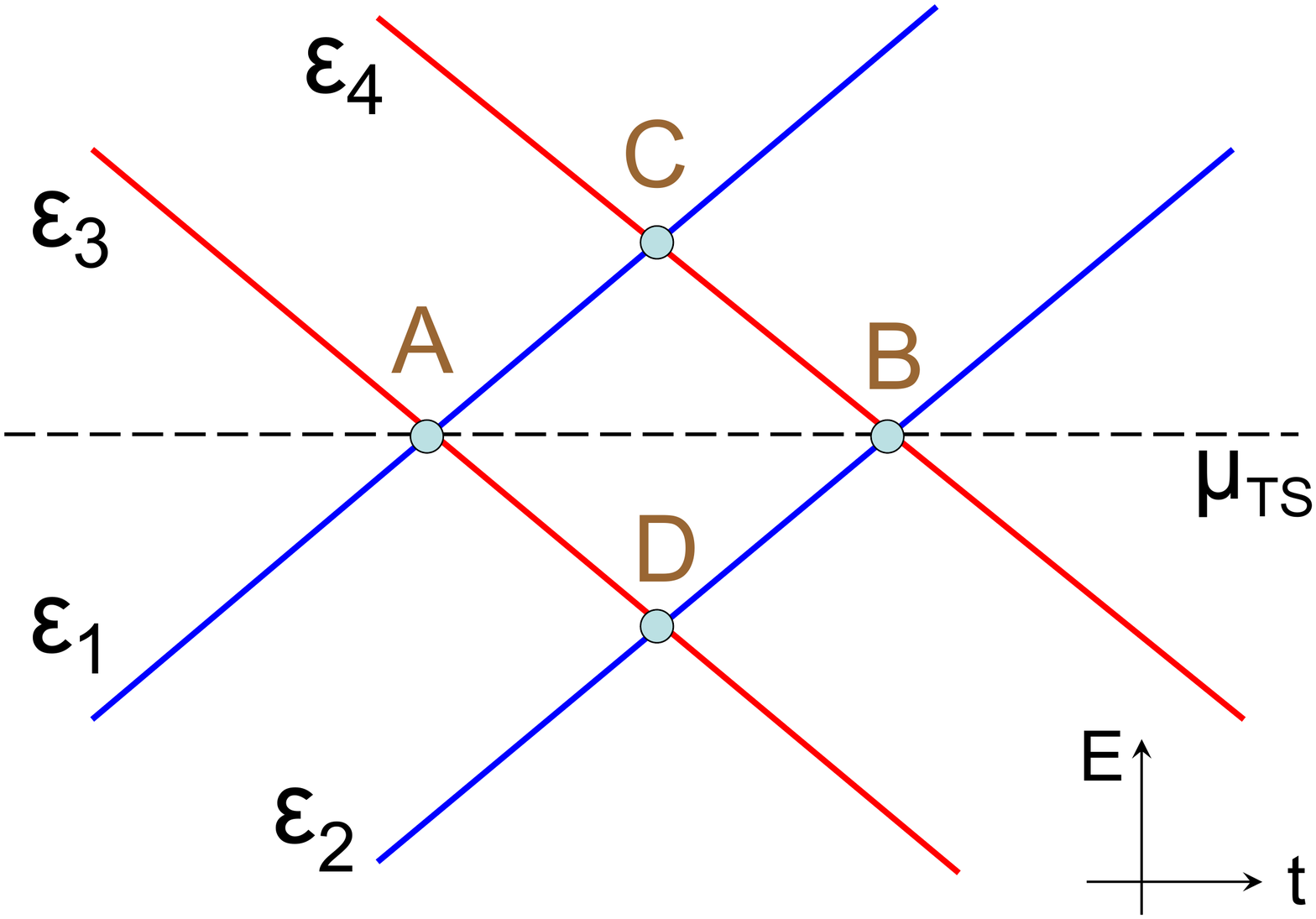}
\caption{\label{fig2} Dot level energy protocol $\{\varepsilon_j(t)\}$ vs time, see Eqs.~\eqref{ham0} and \eqref{protocol1}. }
\end{figure}

As a concrete example, we here consider the schematic setup depicted in Fig.~\ref{fig1}, where a TS island harbors MBSs which are tunnel-coupled to $N=4$ dots as indicated. 
The dot level energy protocol $\{ \varepsilon_j(t) \}$ is illustrated in Fig.~\ref{fig2}.
 It turns out that this is the simplest nontrivial setup where multi-particle interference signals could be observable.
We show that the probability distribution $P[{\bm n}^f]$ is qualitatively different from the corresponding results for setups based on topologically trivial superconductors.  For this comparison, we study a conventional $s$-wave BCS superconductor island without subgap states as well as the case of topologically trivial zero-energy Andreev bound states. 
For the Majorana case, we predict that $P[{\bm n}^f]$ must contain \emph{multi-particle interference} terms for specific ${\bm n}^f$ outcomes that will exhibit robust and quantized $\pi$ phase shifts as one varies a key parameter of the protocol.  
An experimental confirmation of the predicted probability distribution $P[{\bm n}^f]$ would constitute a clear Majorana signature complementary to the information obtained from tunnel spectroscopy \cite{Alicea2012,Lutchyn2017}.   Moreover, we will see that 
the interferometric protocol also allows one to manipulate the Majorana state through 
post-selection.  

Given that a variety of Majorana platforms are explored at the moment,
e.g., semiconductor \cite{Lutchyn2017} or iron-based  \cite{Ironbased1,Ironbased2,Ironbased3} setups, experimental tests should be within reach soon. Once such a platform is available, our protocols impose only modest hardware demands, where the final dot configuration ${\bm n}^f$ can be measured by available 
charge sensing techniques \cite{Bauerle2018,Ensslin}. It stands to reason that similar protocols will allow for many-body interferometric studies of large-scale networks in time-energy space (e.g., using multiple Majorana islands and quantum dot terminals) and/or of systems with richer topological quasiparticles (e.g., parafermionic zero modes).

The remainder of this article is structured as follows. In Sec.~\ref{sec2}, we introduce the low-energy model describing the setup in Fig.~\ref{fig1}, and we discuss the quantum dynamics of the system for the interference protocol sketched in Fig.~\ref{fig2}.  We explore the multi-particle interference phenomena expected for such a Majorana device in Sec.~\ref{sec3}.  Topologically trivial variants of the setup are then studied in Sec.~\ref{sec4}.  Finally, we offer our conclusions and an outlook in Sec.~\ref{sec5}.

\section{Multi-particle interferometry protocols for Majorana devices}\label{sec2}

In this section, we introduce our model for the device in Fig.~\ref{fig1}, see Sec.~\ref{sec2a}. We approach the full multi-particle interference protocol in several steps. First, in Sec.~\ref{sec2b}, we consider the case where only a single dot couples to the TS island and one can map the problem to the standard LZ Hamiltonian.  In Sec.~\ref{sec2c}, we allow for a second dot being coupled to the island and establish a connection to Mach-Zehnder interferometry.  In order to observe HOM-type interferometric signatures, one needs to study the full protocol with all four dots in Fig.~\ref{fig1}.  We address
this case in Sec.~\ref{sec2d}.   The section concludes in Sec.~\ref{sec2e} with a discussion of non-adiabatic transitions at finite energy. In contrast to LZ transitions, such finite-energy transitions are able to couple different dots and thus are needed for generating nontrivial interferometric signatures.

\subsection{Model}\label{sec2a}

We consider a floating TS island with negligible charging energy ($E_C\to 0$), harboring at least four MBSs.  These states are described by Majorana operators $\gamma_j^{}=\gamma_j^\dagger$ 
with the anticommutator algebra $\{\gamma_j,\gamma_k\}=2 \delta_{jk}$.  
We focus on the most interesting case of well-separated MBSs such that the latter represent zero-energy modes.  For the device shown in Fig.~\ref{fig1}, up to
$N=4$ effectively spinless \cite{Alicea2012} single-level quantum dots, described by the fermion annihilation operators $d_j$, are tunnel-coupled with amplitude $\lambda_j$ to individual MBSs.
We choose a gauge where the $\lambda_j$ are real-valued and positive. Moreover, the time-dependent dot level energies $\varepsilon_j(t)$ are taken relative to the TS chemical potential, $\mu_{\rm TS}=0$. 

On energy scales well below the TS pairing gap, above-gap excitations can be neglected and 
the Hamiltonian is  given by  $H(t)= \sum_{j=1}^4 H_j(t)$ with 
\begin{equation}\label{ham0}
 H_j(t)= \varepsilon_j(t) \left( d_j^\dagger d_j^{} -\frac12\right) + \lambda_j \left( d_j^\dagger -d_j^{}\right) \gamma_j.
\end{equation}
 We study protocols $\{\varepsilon_j(t)\}$ of the type shown in Fig.~\ref{fig2}, where the sweep rates $|d \varepsilon_j/dt|$ are always assumed sufficiently low to not excite above-gap quasiparticles. Dot eigenstates are denoted by $|{\bm n}\rangle$ with $\bm n=(n_1n_2n_3n_4)$, where $n_j=0,1$ is the eigenvalue of $d_j^\dagger d_j^{}$.  
 
Given the above model, it is clear that the total number of electrons occupying the four dots may change during the protocol.  In fact, only the total fermion parity of the entire system is conserved.
Equation \eqref{ham0} also shows that the topological degeneracy of 
the uncoupled TS island will be lifted by the tunnel couplings.

\subsection{Landau-Zener transitions} \label{sec2b}

When only dot $1$ is present, i.e., for $\lambda_{2,3,4}=0$, and considering times  $t\approx t_A$,  
where $t_A$ denotes the time at which the dot level crosses the chemical potential of the superconducting island,  $\varepsilon_1(t_A)=\mu_{\rm TS}$ in Fig.~\ref{fig2},
the above setup reduces to the standard LZ problem \cite{LZ1,LZ2,LZ3}.
One can formally show this correspondence by introducing
 composite spin-$1/2$ ladder operators,
\begin{equation}\label{spin12}
    \sigma^{}_{j,+} = d_j^\dagger \gamma_j, \quad \sigma^{}_{j,-} = \sigma_{j,+}^\dagger = \gamma_j d^{}_j,
\end{equation}
resulting in the Pauli operators \cite{foot1}
\begin{eqnarray}
\sigma_{j,x}&=&\sigma_{j,+}+\sigma_{j,-}= (d_j^\dagger-d_j^{})\gamma_j,\nonumber\\
\sigma_{j,z}&=& 2\sigma_{j,+}\sigma_{j,-}-1=2d_j^\dagger d_j^{}-1.
\end{eqnarray}  
For the present case with only $\lambda_1\ne 0$, by writing $\varepsilon_1(t\approx t_A)=\alpha_A(t-t_A)$ with the sweep rate $\alpha_A$, we  obtain
\begin{equation}\label{ham01}
H(t\approx t_A) =  \frac{\alpha_A}{2} (t-t_A)\sigma_{1,z} +\lambda_1 \sigma_{1,x},
\end{equation}
which is identical to the LZ Hamiltonian.
For an arbitrary initial state $|\Phi\rangle$  in the Majorana sector, the ${\bm\sigma}_1$ Pauli operators act in the subspace spanned by the two spinor basis states 
\begin{equation}\label{basisstates}
\left( \begin{array}{c} 1 \\ 0 \end{array} \right)=|n_1=1\rangle\otimes |\Phi\rangle, 
\quad \left( \begin{array}{c} 0 \\ 1 \end{array} \right)=
|n_1=0\rangle\otimes \gamma_1|\Phi\rangle.
\end{equation}   
Incoming ($t=t_A-0^+$) states are then mapped to outgoing  ($t=t_A+0^+$) states 
by the unitary LZ scattering matrix,
\begin{equation}\label{Sta}
S(t_A)= \left( \begin{array}{cc} u_A & \tilde v_A \\ v_A & \tilde u_A  \end{array} \right),
\end{equation}
with  \cite{LZ1,LZ2,LZ3}
\begin{eqnarray}\label{LZsol}
    u_A&=&\tilde u_A=\sqrt{p_A},\quad v_A=-\tilde v^\ast_A = e^{-i\varphi_A}\sqrt{1-p_A},\\ \nonumber
    p_A &= &e^{-2\pi\delta_A},\quad \delta_A= \frac{\lambda_1^2}{2|\alpha_A|},\quad \varphi_A=\frac{\pi}{4}+{\rm arg}\Gamma(1-i\delta_A),
\end{eqnarray}
where $\Gamma$ is the Gamma function, $p_A$ the probability for a successful  LZ transition (unchanged dot occupancy), and $\varphi_A$ the LZ phase shift picked up otherwise.
From the above discussion, we also observe that the scattered entities in our interference protocol will not just be electrons. Scattering processes instead involve composite spin-1/2 particles obtained by combining electrons and Majorana particles, cf.~Eq.~\eqref{spin12}.

\subsection{Mach-Zehnder interferometer}\label{sec2c}

To approach the full protocol shown in Fig.~\ref{fig2}, in a next step we shall allow for a finite coupling of dot $4$ to the island ($\lambda_4\ne 0$) as well.  We still  
keep $\lambda_{2,3}=0$ and start from the initial product state 
$|\Psi(t_i)\rangle=|10\rangle \otimes |\Phi\rangle$, where dot 1 is initially occupied and dot 4 is empty. 
All in all, the dots are initially occupied by a single electron and the setup is reminiscent of a Mach-Zehnder interferometer \cite{Hariharan2007,LZ3}. 

Besides the LZ transition at $t_A$, we now encounter a second LZ transition near the time $t_B$ with $\varepsilon_4(t_B)=\mu_{\rm TS}$. 
In addition,  a qualitatively new type of transition appears near the time $t_C$ in Fig.~\ref{fig2}, where a finite-energy level crossing  occurs,
\begin{equation}\label{tcdef}
\varepsilon_1(t_C)=\varepsilon_4(t_C)\equiv \varepsilon_C.
\end{equation} 
For $t\approx t_C$, non-adiabatic transitions due to \emph{elastic cotunneling} across the island can take place, see Sec.~\ref{sec2e} for details. 
Throughout, we assume that the condition \cite{Mullen1989}
\begin{equation}\label{mullen}
 {\rm max}\left(\lambda_j,\sqrt{\alpha_A}\right)\ll \varepsilon_C
 \end{equation}
 is satisfied.  In practice, its validity can be ensured by increasing the time difference $t_B-t_A$.  If Eq.~\eqref{mullen} holds, we have well-separated non-adiabatic regions near $t_A, t_B$, and $t_C$,  where the respective transitions are described by scattering matrices $S(t_{B,C})$ of similar form as in Eq.~\eqref{Sta}.
  
The final dot occupation probabilities, $P[{\bm n}^f]={\cal P}_{n_1n_4}$, observed at time $t_f>t_B$ then readily follow as
\begin{eqnarray}\nonumber
{\cal P}_{00} &=& \left| u_A v_C v_B + e^{i\chi} v_A u_B \right|^2,  
 \\ \label{mzi}
{\cal P}_{01} &=& \left| u_A v_C u_B + e^{i\chi} v_A \tilde v_B \right|^2, \\ \nonumber
{\cal P}_{10} &=& |u_Au_Cu_B|^2,
\quad {\cal P}_{11} = |u_A u_C\tilde v_B|^2,
\end{eqnarray}
where we define the dynamical phase 
\begin{equation}\label{dynphase}
\chi=\chi_4(t_B,t_C) + \chi_1(t_C,t_A),\quad  \chi_j(t,t')=\int_{t'}^t d\tau \varepsilon_j(\tau),
\end{equation}
 which is picked up during the adiabatic stages of the time evolution. 
 
 Clearly, Eq.~\eqref{mzi} shows that the probabilities ${\cal P}_{n_1=0, n_4}$ contain an interference term causing 
 Landau-Zener-St\"uckelberg oscillations  \cite{LZ3,Shimshoni1991}. These oscillations are similar to those previously predicted for other MBS systems \cite{Flensberg2011,Scheurer2013,Huang2015,Knapp2016,Wang2018,Khlebnikov2018,Bauer2018}.
 The above results also show that
in order to observe HOM-type multi-particle interference phenomena, one needs to go beyond a setup with only $N=2$ terminals.

\subsection{HOM-type protocol}\label{sec2d}

We next consider the full protocol in Fig.~\ref{fig2} with all four tunneling amplitudes $\lambda_j\ne 0$. Our protocol starts from a general product state, 
\begin{equation}
|\Psi(t_i)\rangle= |{\bm n}^i\rangle\otimes |\Phi\rangle,
\end{equation}
again with an arbitrary initial state $|\Phi\rangle$ in the Majorana sector.  
We first observe that Eqs.~\eqref{ham0} and \eqref{basisstates} imply that every dot occupancy change, $n_j\to 1-n_j$, must be accompanied by a transformation of the Majorana state,  $|\Phi\rangle\to \gamma_j |\Phi \rangle$. As a consequence of this one-to-one correspondence, the full quantum state dynamics must be of the generic form
\begin{equation}\label{dynamics1}
|\Psi(t>t_i)  \rangle =\sum_{\{{\bm n}\}} |{\bm n}\rangle\otimes {\cal A}_{\bm n,\bm n^i}[\gamma] |\Phi\rangle, 
\end{equation}
with the operators
\begin{equation}
{\cal A}_{\bm n, \bm n^i}[\gamma] =
C_{\bm n,\bm n^i} \prod_{j=1}^4 \gamma_j^{\left|n_j-n_j^i\right|},
\end{equation}
 with $\gamma_j^0=1$ and complex coefficients $C_{\bm n, \bm n^i}$ which depend on the precise form of the protocol.
Using Eq.~\eqref{dynamics1}, the final dot occupation probability distribution follows as
\begin{equation}\label{finalpro}
    P[\bm n^f] = \left\langle \Phi\right| {\cal A}_{\bm n^f,\bm n^i}^\dagger
    {\cal A}^{}_{\bm n^f, \bm n^i}\left|\Phi\right\rangle    =\left |C_{\bm n^f,\bm n^i}\right|^2,
\end{equation}
which is completely \emph{independent} of the initial Majorana state $|\Phi\rangle$.

Nonetheless, measuring a specific outcome ${\bm n}^f$ implies that $|\Phi\rangle$ has been changed according to Eq.~\eqref{dynamics1}.  Our protocol thus offers a way to facilitate Majorana state manipulation by post-selection.
While only Clifford operations can be realized by the above protocol, we note that arbitrary phase gates could be accessed when adding tunnel couplings (reference arms) between selected pairs of quantum dots, cf.~Refs.~\cite{Plugge2017,Karzig2017}. 

To simplify the algebra, from now on we shall write the protocol in Fig.~\ref{fig2} in the specific form
\begin{equation}\label{protocol1}
\varepsilon_1(t)=-\varepsilon_3(t)=\alpha(t-t_A),
\quad
\varepsilon_2(t)=-\varepsilon_4(t)=\alpha(t-t_B),
\end{equation}
with identical sweep rate $\alpha$ for all quantum dots. 
Since different $H_j$ terms in Eq.~\eqref{ham0}  commute, the precise order of both LZ transitions at $t=t_A$ (where $\varepsilon_1$ and $\varepsilon_3$ cross 
$\mu_{\rm TS}$, respectively) is irrelevant. It is thus safe to assume that they 
happen simultaneously. 
The same argument applies to the two LZ transitions at $t_B$.   
We emphasize that the LZ transitions at $t_{A,B}$ do not introduce correlations between different dots. In fact, they play a similar role as the beam splitter in the standard HOM setup.

In addition to the LZ transitions at $t_{A,B}$,  non-adiabatic finite-energy transitions occur at the times $t_C$ and $t_D$ in Fig.~\ref{fig2}.  Using the specific protocol in  Eq.~\eqref{protocol1}, we obtain
\begin{equation}
t_C=t_D=(t_A+t_B)/2,
\end{equation}
resulting in 
\begin{equation}
\varepsilon_C=\alpha(t_B-t_A)/2, \quad \varepsilon_D=-\varepsilon_C.
\end{equation}
As shown in Sec.~\ref{sec2e}, finite-energy transitions generate correlations between incoming particles and thereby can be responsible for a nontrivial interference pattern. 

We note that the assumptions behind Eq.~\eqref{protocol1} are less restrictive than they may appear at first sight.  
Indeed, the precise form of the protocol during the adiabatic stages of the evolution is irrelevant.
Similarly, by allowing for a non-zero time difference $\tau=t_C-t_D$, one can implement a delay time between incoming particles.  By increasing $\tau$, their correlations could be effectively switched off as in the HOM setup  \cite{Hong1987}.  However, we focus on the $\tau=0$ case below.

\subsection{Finite-energy transitions}\label{sec2e}

Consider now the vicinity of a  finite-energy crossing,  $t\approx t_{X=C,D}$, see Fig.~\ref{fig2}.
Averaging over fast oscillations corresponding to transition energies of order $\varepsilon_C$,  Eq.~\eqref{ham01} yields effective exchange interactions between pairs $j\ne k$ of the effective spin-1/2 operators in Eq.~\eqref{spin12}. 
With $\nu,\nu'=\pm 1$, these interactions have the form
 \begin{equation}\label{exchange}
 H_{\rm ex} (t\approx t_{X}) = w_{jk} \varepsilon_C \left( \sigma_{j,\nu}\sigma_{k,\nu'}+ {\rm h.c.}\right)  
\end{equation}
with the dimensionless exchange couplings
\begin{equation}\label{exchcoup}
w_{jk} \simeq \frac{ \lambda_j\lambda_k}{\varepsilon_C^2 }\ll 1.
\end{equation}
In physical terms, $\nu=-\nu'= \pm 1$ in Eq.~\eqref{exchange} describes cotunneling processes where an electron is transferred between dots $j$ and $k$ across the island under the condition $\varepsilon_j(t_X)=\varepsilon_k(t_X)$.  Note that $H_{\rm ex}$ in Eq.~\eqref{exchange} then acts 
as $d_j^\dagger d_k$ or $d_k^\dagger d_j$ in the dot Hilbert space.
Terms with $\nu=\nu'$ instead describe crossed Andreev reflection (CAR) processes, which are possible for $\varepsilon_j(t_X)+\varepsilon_k(t_X)= \mu_{\rm TS}$ and correspond to two-electron absorption or emission by the TS condensate, $H_{\rm ex}\propto d_j^\dagger d_k^\dagger + d_k d_j$.  Finite-energy transitions therefore introduce correlations between incoming 
particles.

\section{Multi-particle interferometry in the time-energy domain}\label{sec3}

In this section, we start by discussing
 general multi-particle interferometric features for Majorana devices, see Sec.~\ref{sec3a}.
This topologically nontrivial case can be identified by a characteristic $\pi$ phase shift of the interferometric signal contained in $P[{\bm n}^f]$ for certain final dot configurations, see Sec.~\ref{sec3b}.

\subsection{General principles}\label{sec3a}

The final state has the structure
\begin{eqnarray}\nonumber
&&|\Psi(t_f)\rangle = \sum_{\{\bm n_A,\bm n_C, \bm n_B\}} \Omega_{\bm n_B}(t_f,t_B) {\cal S}(t_B) \Omega_{\bm n_C}(t_B,t_C) \\ 
&&\times {\cal S}(t_C) \Omega_{\bm n_A}(t_C,t_A) {\cal S}(t_A) \Omega_{\bm n^i}(t_A,t_i)|\Psi(t_i)\rangle,\label{schematic}
\end{eqnarray}
where 
\begin{equation}
    \Omega_{\bm n}(t,t')=e^{-i\sum_j\chi_j(t,t')n_j}
\end{equation}
contains dynamical phase factors, see Eq.~\eqref{dynphase}.
The operators  ${\cal S}(t_{A,B,C})$ in Eq.~\eqref{schematic} describe non-adiabatic transitions at the respective times. In particular,
${\cal S}(t_{A})$ is a product of the two uncorrelated LZ scattering matrices $S(t_A)$ for $H_1$ and $H_3$, cf.~Eq.~\eqref{ham0}.
Similarly, ${\cal S}(t_B)$ follows as product of the LZ matrices $S(t_B)$ for $H_2$ and $H_4$.  

The operator ${\cal S}(t_C)$ encodes correlations due to cotunneling and/or CAR processes near $t_C=t_D$, see Sec.~\ref{sec2e}.
By expanding in the small exchange couplings $w_{jk}\ll 1$,  see Eq.~\eqref{exchcoup}, we find that ${\cal S}(t_C)$ acts on the state $|{\bm n}\rangle\otimes |\Phi\rangle$ as
\begin{equation}\label{stc}
    {\cal S}(t_C) = \mathtt{1} + \sum_{j<k} w_{jk} (-1)^{n_j} \xi_{\bm n} L_{jk} + {\cal O}\left(w_{jk}^2\right), 
\end{equation}
where $\xi_{\bm n}=1$ (2) for even (odd) values of $\sum_j n_j$ and $L_{13}=L_{24}=0$. 
In Eq.~\eqref{stc}, only low-energy states with energy difference well below $\varepsilon_C$ have been kept, such that $n_1-n_2-n_3+n_4$ is conserved at the non-adiabatic transition.  Cotunneling processes are encoded by
\begin{equation}\label{cotu}
    L_{14} = \delta_{n_1,1-n_4} \left(d_1^\dagger - d^{}_1\right)\gamma_1 \left(d_4^\dagger - d^{}_4\right)\gamma_4 , 
\end{equation}
and likewise for $L_{23}$, and CAR processes are contained in 
\begin{equation}\label{CAR}
    L_{12} = \delta_{n_1,n_2} \left(d_1^\dagger - d^{}_1\right)\gamma_1 \left(d_2^\dagger - d^{}_2\right)\gamma_2,
\end{equation}
where (up to an overall sign change) $L_{34}$ follows analogously. We note in passing that
Eqs.~\eqref{cotu} and \eqref{CAR} directly correspond to the exchange processes specified in Eq.~\eqref{exchange}.

From Eq.~\eqref{schematic}, with $\bar {\bm n}$ denoting the particle-hole reversed configuration ($\bar n_j= 1-n_j$), we find that 
the probability distribution satisfies a general symmetry relation, 
\begin{equation}\label{phs}
    P\left[\bm n^f;\bm n^i;\chi\right]= P\left[\bar {\bm n}^f;\bar {\bm n}^i ;\chi+\pi\right],
\end{equation}
where we explicitly include the dependence on the initial dot configuration $\bm n^i$ and on the dynamical phase 
\begin{equation}\label{chidef}
\chi=\frac{ \alpha}{4}(t_B-t_A)^2.
\end{equation}.

\begin{figure}
\centering
\includegraphics[width=\columnwidth]{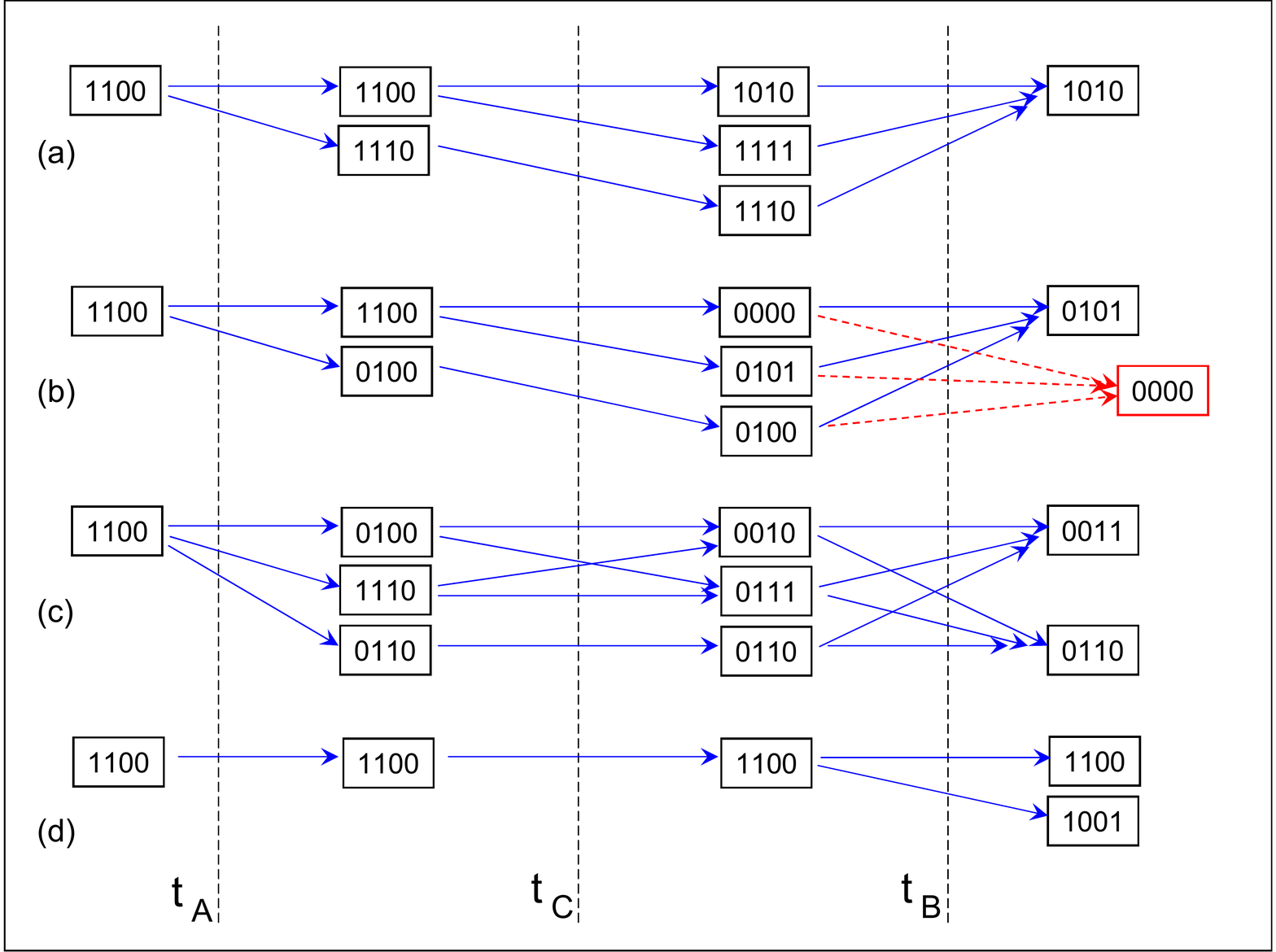}
\caption{\label{fig3} Possible trajectories for the dot configurations $\bm n=(n_1n_2n_3n_4)$ under the protocol in Fig.~\ref{fig2} when starting from the two-electron configuration 
$\bm n^i=(1100)$.   During the adiabatic stages, $\bm n(t)$ does not change. 
Non-adiabatic transitions (blue arrows) at $t\approx t_{A,C,B}$ can generate 
interfering trajectories for certain final configurations $\bm n^f$ at $t_f>t_B$.  
For clarity,  transitions into states with odd $\sum_j n_j^f$ are not shown. 
}
\end{figure}

\begin{figure}
\includegraphics[width=\columnwidth]{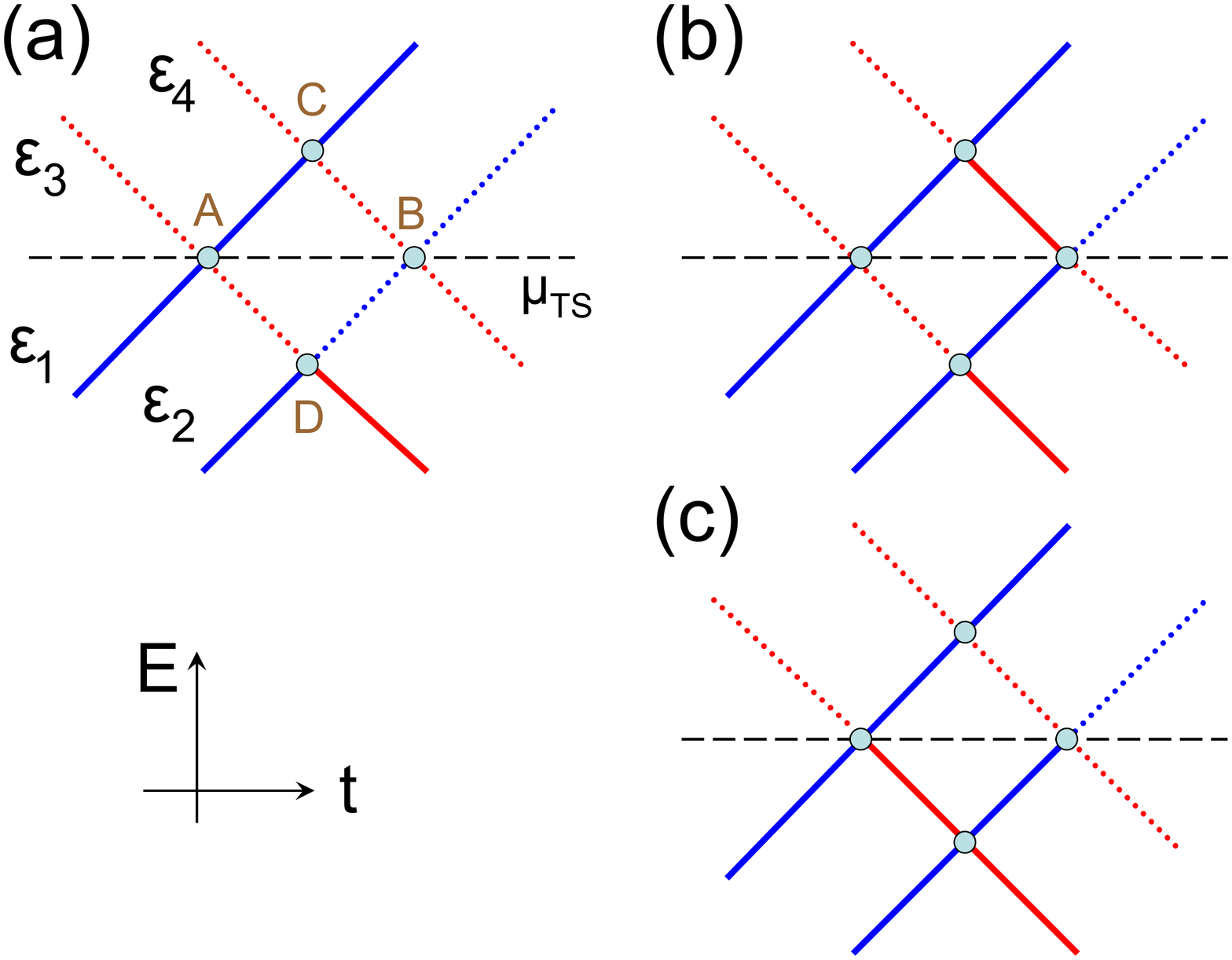}
\caption{\label{fig4}
Possible trajectories for the transition ${\bm n}^i=(1100)\to {\bm n}^f= (1010)$ in the time-energy domain, see Figs.~\ref{fig2} and \ref{fig3}(a).  Solid (dotted) lines correspond to occupied (empty) dot levels. The initially occupied dots 1 and 2 are shown in blue, the initially empty dots 3 and 4 in red.
(a) Cotunneling trajectory. (b) CAR trajectory. (c) Reference trajectory without high-energy transitions.  The interference of those three trajectories can be observed by monitoring the  
occupation probability $P[{\bm n}^f]$, see main text.
}
\end{figure}

Figure~\ref{fig3} summarizes the possible multi-particle trajectories starting from ${\bm n}^i=(1100)$  and ending at some configuration ${\bm n}^f$.    Let us start with the diagram in Fig.~\ref{fig3}(a). Without a non-adiabatic finite-energy transition at $t_C$, one has the uncorrelated reference path ${\bm n}^i\to (1110) \to {\bm n}^f=(1010)$.  This specific multi-particle trajectory is also illustrated in the time-energy diagram of Fig.~\ref{fig4}(c).
Including a finite-energy transition at $t_C$,  two additional trajectories are generated. 
The first trajectory involves a cotunneling event, with ${\bm n}^i\to (1010)$ at $t_C<t<t_B$.
The corresponding time-energy diagram is shown in Fig.~\ref{fig4}(a).  The second trajectory is generated by a CAR process, where ${\bm n}^i\to (1111)$ at $t_C<t<t_B$, see Fig.~\ref{fig4}(b).
Through the LZ transition at $t\approx t_B$, both these trajectories may transit into the selected final state ${\bm n}^f$. 
The interference of those three trajectories leaves clear signatures in $P[\bm n^f]$ as shown in Sec.~\ref{sec3b} below.
  
More generally, finite-energy transitions can generate a multitude of trajectories with relative weight $\sim w_{jk}$ on top of a reference path, which may then interfere with each other.
On the other hand, for ${\bm n}^f=(1100)$ or $(1001)$, we observe from Fig.~\ref{fig3}(d) that no interference is possible  since only a single multi-particle trajectory exists.

\subsection{Interference signature for Majorana states}\label{sec3b}

To further simplify the notation, we next assume that all tunnel couplings are equal, $\lambda_j=\lambda$.    
For every LZ transition, we thus have the same success probability $p=e^{-\pi\lambda^2/\alpha}$ and the same phase shift $\varphi$. Moreover, cotunneling and CAR transitions at $t\approx t_C$  involve just one  parameter, $w=\lambda^2/\varepsilon_C^2$.  

For $\bm n^i=(1100)$, see  Fig.~\ref{fig3}, the probabilities 
$P[\bm n^f]={\cal P}_{n_1n_2n_3n_4}$ with even $\sum_j n_j$ then take the following form.
As expected from Fig.~\ref{fig3}(d), ${\cal P}_{1100} = p^4$ and ${\cal P}_{1001} = p^2 (1-p)^2$ do not contain interference terms.
However, all other probabilities for the outcomes in Fig.~\ref{fig3} exhibit 
oscillatory St\"uckelberg-like interference terms $\sim \cos\eta$ with the 
phase 
\begin{equation}\label{etadef}
 \eta= \chi - 2\varphi.
\end{equation}
In particular, we find
\begin{equation} \label{Ptwo1}
 {\cal P}_{1010} = {\cal P}_{0101} = {\cal P}_{0000} = p^2 \left|1-p-w(2p-1)e^{i\eta}\right|^2,
\end{equation}
and
\begin{eqnarray}
{\cal P}_{0011} &=& (1-p)^2 \left| 1-p - 4 w p e^{i \eta} \right|^2, \nonumber\\
{\cal P}_{0110} &=& p^2(1-p)^2 \left| 1 + 4 w e^{i \eta} \right|^2.\label{Ptwo2}
\end{eqnarray}

Remarkably,
for the three  measurement outcomes in Eq.~\eqref{Ptwo1}, we encounter an interference signal $\propto w(2p-1)\cos\eta$.  
By changing the LZ probability across the critical value $p=p_c=1/2$, the interference pattern  in $P[{\bm n}^f]$  effectively acquires a phase shift of $\pi$.  Such phase shifts 
occur only for the specific outcomes in Eq.~\eqref{Ptwo1} and can be traced back to the interplay of trajectories with a cotunneling and a CAR transition. 
For this reason, one  needs at least $N=4$ terminals, cf.~Fig.~\ref{fig1}, in order to observe this effect.

In practice, one could either change the tunnel couplings $\lambda$ to detect this phase shift, or change the sweep rate $\alpha$ with the dynamical phase $\chi$ kept constant by adjusting $t_B-t_A$.  Writing $t_B-t_A=q\lambda/\alpha$ with a factor $q\gg 1$, we have $\chi=-\frac{q^2}{4\pi} \ln p$, resulting in $\chi\sim 2\pi$ for $p\sim 1/2$ and $q\sim 10$. 
We show in Sec.~\ref{sec4} that the $\pi$ shift is not expected for topologically trivial settings.  In this sense, its observation could represent a clear signature for Majorana states.
Finally, let us note that the interference phase extracted from $P[{\bm n}^f]$ will in general include the dynamical phase $\chi$, LZ phase shifts $\varphi$, and a statistical phase \cite{Snizhko2019a,Snizhko2019b} obtained by averaging geometric phases over many realizations.

\section{Topologically trivial cases} \label{sec4}

Above we have argued that features like the $\pi$ shift in the interference pattern represent a characteristic signature for Majorana states.  To further support this claim,
we have carried out a similar analysis for two topologically trivial variants of the above setup.
The first variant has no subgap states at all, see Sec.~\ref{sec4a}, and the second case arises when 
 MBSs are replaced by zero-energy Andreev bound states, see Sec.~\ref{sec4b}. As we show below, both cases can 
easily be distinguished from the true Majorana setup where each dot couples only to a single MBS.  
Finally, for a floating Majorana island with large charging energy $E_C$, see Refs.~\cite{Plugge2017,Karzig2017}, the setup in Fig.~\ref{fig1} will be analyzed by similar methods in Sec.~\ref{sec4c}.

\subsection{No subgap states}\label{sec4a}

In the absence of subgap states, we use the Hamiltonian 
\begin{equation}\label{tt:H}
H(t) = \sum_{j = 1}^4 \left[\varepsilon_j(t) \left( d_j^\dagger d_j - \frac12 \right) + V_j \right] + \sum_\nu E_\nu \gamma_\nu^\dagger \gamma_\nu,
\end{equation}
where the island is represented by continuum quasi-particles with energies $E_\nu\ge \Delta$, with the corresponding
fermionic operator $\gamma_\nu$ for quantum numbers $\nu$.
The tunnel contacts are described by
\begin{equation}
V_j = d_j^\dagger \sum_\nu \left( a_{j \nu} \gamma_\nu + b_{j \nu} \gamma_\nu^\dagger \right) + {\rm h.c.},
\end{equation}
with complex-valued normal and anomalous tunneling amplitudes $a_{j \nu}$ and $b_{j \nu}$, respectively.
We again assume a large superconducting gap, $\Delta \gg \max \left( | \varepsilon_C|, \lambda_j \right)$, and
an initial product state, $|\Psi(t_i)\rangle = |{\bm n}^i \rangle \otimes |\Phi_0\rangle$, where 
$|\Phi_0\rangle$ denotes the BCS ground state of the island with $\gamma_\nu |\Phi_0\rangle = 0$ for all $\nu$.

Using a Schrieffer-Wolff transformation, we next project the Hamiltonian \eqref{tt:H} to the low-energy subspace (valid on subgap scales) 
for each of the three non-adiabatic regions at $t \approx t_{A,B,C}$. 
The LZ transitions at $t_{A,B}$ are governed by an effective two-level Hamiltonian,
\begin{eqnarray}\label{tt:HtAB}
H(t \approx t_{A,B}) &=& \frac{\varepsilon_j(t)}{2} \left( d_j^\dagger d_j^{} - d_k^\dagger d_k^{} \right) \\ \nonumber 
&+&\left( \lambda_{jk} d_j^\dagger d_k^{} + \Delta_{jk} d_j^\dagger d_k^\dagger + {\rm h.c.} \right),
\end{eqnarray}
where 
\begin{eqnarray}\nonumber
\lambda_{jk} &=& - \sum_\nu  \left( a_{j \nu}^{} a_{k \nu}^\ast -  b^{}_{j \nu} b_{k \nu}^\ast \right)/E_\nu,\\ 
\Delta_{jk} &=& \sum_\nu \left( a_{j \nu} b_{k \nu} -  b_{j \nu} a_{k \nu} \right)/E_\nu,
\end{eqnarray}
with $(j,k) = (1,3)$ and $(2,4)$ for $t \approx t_A$ and $t_B$, respectively. 
In contrast to the Majorana case, the total fermion number parity of the dots is now preserved in the non-adiabatic regions near $t_{A,B}$.
In fact, transitions to above-gap quasiparticle states average out on timescales above $1/\Delta$.

For the non-adiabatic transitions at $t\approx t_C$,  we take into account only processes  relevant
on timescales of order $1/\varepsilon_C$.
To lowest order in $\lambda_{jk}$ and $\Delta_{jk}$, all non-vanishing matrix elements follow from
the relations
\begin{eqnarray}\nonumber
| 0 ,0, n_3, n_4 \rangle &\rightarrow& \Delta_{12} | 1,1, n_3, n_4 \rangle,\\ \nonumber
| n_1, n_2, 0 ,0 \rangle &\rightarrow& \Delta_{34} | n_1, n_2,1,1 \rangle, \\ \nonumber
| 0, n_2, 0 ,n_4 \rangle &\rightarrow& (-1)^{n_2} \Delta_{13} | 1 ,n_2, 1, n_4 \rangle,\\
| n_1 ,0, n_3, 0 \rangle &\rightarrow& (-1)^{n_3} \Delta_{24} | n_1, 1 ,n_3, 1 \rangle, \\ \nonumber
| 0, n_2, n_3, 1 \rangle &\rightarrow & (-1)^{n_2+n_3} \lambda_{14} | 1 ,n_2, n_3, 0 \rangle,\\ \nonumber
| n_1 ,0 ,1, n_4 \rangle &\rightarrow & \lambda_{23} | n_1 ,1 ,0 ,n_4 \rangle,
\end{eqnarray}
plus the conjugate processes.

\begin{figure}
\includegraphics[width=\columnwidth]{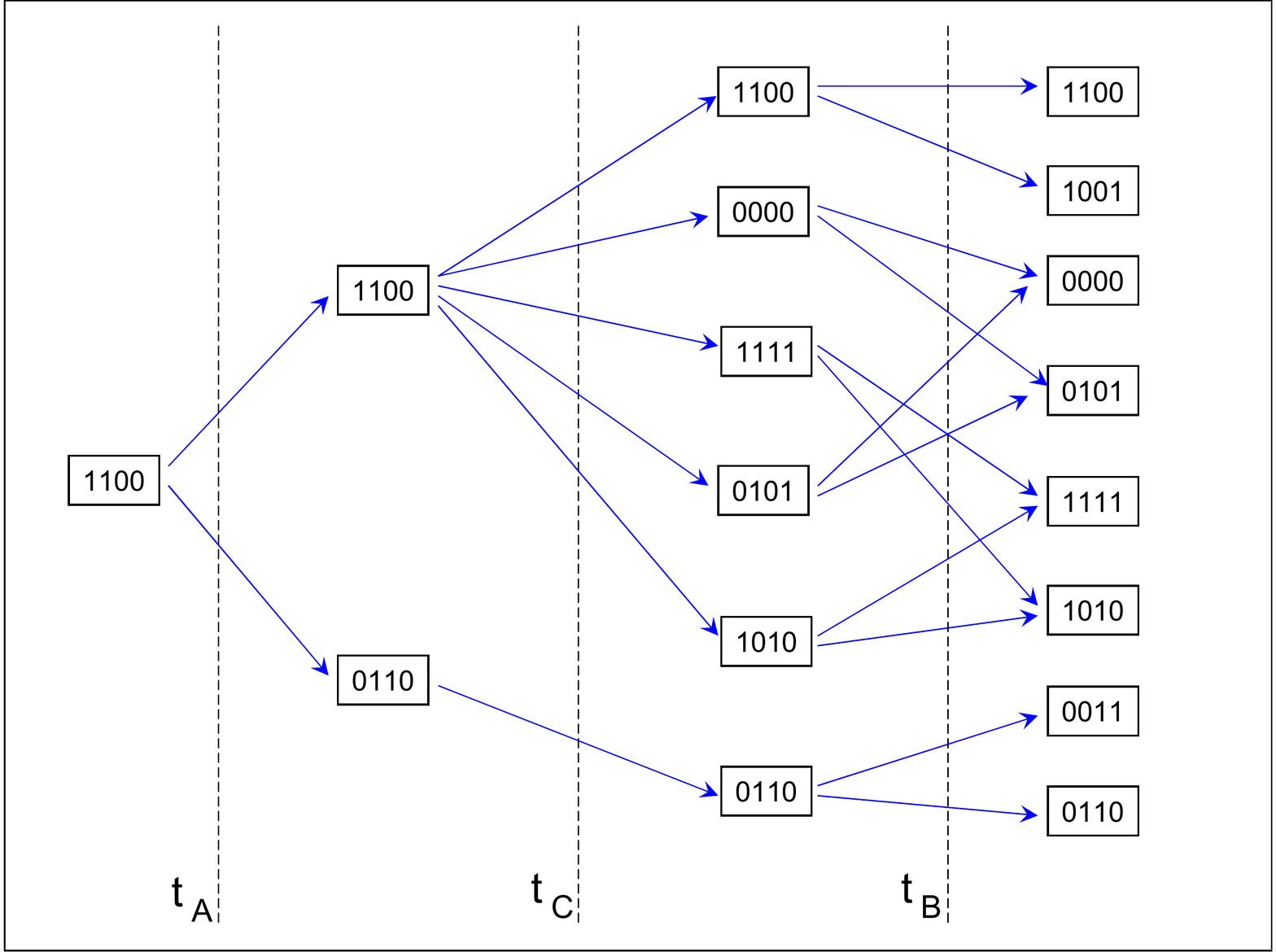}
\caption{\label{fig5}
Same as in Fig.~\ref{fig3} but for an $s$-wave BCS island without subgap states.
Starting from an initial state with $\bm n^i=(1100)$, only parity-conserving transitions are allowed.
}
\end{figure}

Figure~\ref{fig5} shows the possible dot occupation number trajectories
starting from the initial configuration $\bm n^i=(1100)$.
Although many probabilities $P[{\bm n}^f]$ involve interfering multi-particle trajectories,
see Fig.~\ref{fig5}, we find that none of them depends on the dynamical phase shift $\chi$ 
in Eq.~\eqref{chidef}. The interference pattern is therefore not adjustable and hence 
strongly differs from the Majorana case. 

We note that
the model in Eq.~\eqref{tt:H} may also be useful for analyzing effects caused by above-gap quasiparticles (e.g., at finite temperature) in the Majorana device studied in Secs.~\ref{sec2} and \ref{sec3}. While we have not carried out a detailed analysis, we anticipate that $P[\bm n^f]$ will then weakly depend on the initial Majorana state $|\Phi\rangle$.

\subsection{Andreev bound states}\label{sec4b}

As second example for a topologically trivial device,
we next consider a setup where the dots are connected to the superconducting island through  
fermionic operators $f_j$ representing zero-energy Andreev (instead of Majorana) bound states. 
In this part, we again neglect above-gap quasiparticles and use the model 
\begin{equation}\label{absmod}
H=\sum_j H_j(t),\quad H_j(t)=\sum_j\left[\varepsilon_j(t) \left(d_j^\dagger d_j^{}-\frac12 \right)+V_j\right],
\end{equation}
 where tunneling is described by 
\begin{equation}\label{tt2:H}
V_j = d_j^\dagger \left( \lambda_j f_j + \delta_j f_j^\dagger \right) + {\rm h.c.}
\end{equation}
We choose a gauge with real-valued tunnel couplings $\lambda_j \ge 0$, and also include complex-valued anomalous tunnel couplings $\delta_j$. 
The Majorana case is recovered for $\lambda_j = \delta_j$.

The Hilbert space of the complete system is spanned by the Fock states (with $n_j,m_j=0,1$)
\begin{equation}
| n_1 m_1,  n_2 m_2, n_3 m_3, n_4 m_4 \rangle =
\prod_{j=1}^4 \left( d_j^\dagger \right)^{n_j} \left( f_j^\dagger \right)^{m_j} |{\rm vac} \rangle,
\end{equation}
with the empty state $|{\rm vac}\rangle$ of the full system. The non-adiabatic transitions at $t\approx t_{A,B,C}$ are then described by similar expressions for the multi-particle scattering operators ${\cal S}(t_{A,B,C})$ as in Sec.~\ref{sec3a}. However, for $\lambda_j\ne \delta_j$,
the corresponding scattering amplitudes, i.e., $u$, $v$ and $w$, will now depend on the 
parities $(-1)^{n_j + m_j}$ of the respective incoming states.

Introducing parity-dependent LZ amplitudes,
$\left( u_j^{(e/o)}, v_{j, n_j}^{(e/o)} \right)$, with
\begin{equation}
u_j^{(e/o)} = u_j^{(e/o) \ast},\quad v_{j,1}^{(e/o)} = -v_{j, 0}^{(e/o) \ast} \equiv v_j^{(e/o)},
\end{equation}
and parity-dependent cotunneling amplitudes ($j<k$),
\begin{equation}
    w_{jk, n_j n_k}^{(e/o , e/o)} = \frac{\lambda^{(e/o)}_{j, n_j} \lambda^{(e/o)}_{k, n_k} }{ \varepsilon_C^2},
\end{equation}
where we use the notation
\begin{equation}
\lambda^{(o)}_{j, 0/1} = \lambda_j \equiv \lambda^{(o)}_j,\quad  
\lambda^{(e)}_{j,0} = \lambda^{(e) \ast}_{j,1} = \delta_j \equiv \lambda^{(e)}_j,
\end{equation}
we arrive (for instance) at the relations
\begin{widetext}
\begin{eqnarray} \nonumber
| 10, 10, 00, 00 \rangle \Big|_{t = t_A} &\rightarrow& u_1^{(o)} u_3^{(e)} | 10, 10, 00, 00 \rangle +
v_1^{(o)} v_3^{(e)} | 01, 10, 11, 00 \rangle +
v_1^{(o)} u_3^{(e)} | 01, 10, 00, 00 \rangle +
u_1^{(o)} v_3^{(e)} | 10, 10, 11, 00 \rangle,\\ \nonumber
| 10, 10, 00, 00 \rangle \Big|_{t = t_C} &\rightarrow & | 10, 10, 00, 00 \rangle
- w_{12,11}^{(oo)} | 01, 01, 00, 00  \rangle
- w_{34,00}^{(ee)} | 10, 10, 11, 11 \rangle \\ && \qquad - w_{14,10}^{(oe)} | 01, 10, 00, 11 \rangle
- w_{23,10}^{(oe)} | 10, 01, 11, 00 \rangle.
\end{eqnarray}
\end{widetext}
The possible dot configuration trajectories starting from ${\bf n}^i=(1100)$ are  then 
again described by Fig.~\ref{fig3}. 
However, the corresponding expressions for $P[\bm n^f]$ now contain \emph{parity-resolved} scattering amplitudes,
and hence $P[\bm n^f]$ will depend on the initial island state $|\Phi_{\rm S}\rangle$.

For quantitive results, we consider equal couplings, 
$\lambda_j = \lambda > 0$ and $\delta_j = \delta = \delta^\ast$.
We then find
\begin{eqnarray}
&& u_{1,2,3,4}^{(e/o)} \equiv u_{e/o},\quad v_{1,2}^{(e/o)} = -v_{3,4}^{(e/o) \ast} \equiv v_{e/o},\\ \nonumber
&& w_{oo} = \frac{\lambda^2}{\varepsilon_C^2},\quad w_{ee} =\frac {\delta^2}{\varepsilon_C^2},\quad w_{oe} = w_{eo} = 
\frac{\lambda \delta}{\varepsilon_C^2}.
\end{eqnarray}
Let us give a few  examples. 
For $|\Psi(t_i)\rangle = |\bm n^i \rangle \otimes |\Phi_{\rm S}\rangle$, with
$\bm n^i=(1010)$ and 
$|\Phi_{\rm S}\rangle = |mmmm \rangle$, we obtain the probabilities $P[\bm n^f]={\cal P}_{n_1n_2n_3n_4}$ in closed form.
We here specify only two of them for $m=0$,
\begin{eqnarray}
{\cal P}_{1100} &=& \left|
u_o v^\ast_o v_e^\ast u_e + w_{oe} \left( u^{2}_o - |v_o|^2  \right) u^{2}_e e^{i \chi} \right|^2,\\
\nonumber
{\cal P}_{1001} &=& \left| \label{finie}
u_o v^\ast_o  u_e v_e - w_{oe} \left( u^{2}_o - |v_o|^2  \right) v^{2}_e e^{i \chi} \right|^2.
\end{eqnarray}
The corresponding results for $m=1$ follow by interchanging the parity indices, $e \leftrightarrow o$, and the
extension to general states $|\Phi_{\rm S} \rangle= |m_1 m_2 m_3 m_4 \rangle$  
is also straightforward. 
In particular, for $\lambda=\delta$ (implying parity-independent $u, v$ and $w$),
the Majorana results of Sec.~\ref{sec3} are recovered for arbitrary initial state $|\Phi_{\rm S}\rangle$.

While the $\pi$ phase shift described in Sec.~\ref{sec3b} can here also occur for specific initial states, cf.~Eq.~\eqref{finie},
 $P[{\bm n}^f]$ now depends on the initial state $|\Phi_{\rm S}\rangle$. 
When repeating the measurement of the final dot configuration many times in order to obtain the probability distribution $P[\bm n^f]$, the $\pi$ phase shift is thus completely washed out.
(The same conclusion applies when quasiparticle poisoning is important.)
By contrast, when individual Majorana states are coupled to the respective dot, $P[\bm n^f]$ is
 independent of $|\Phi\rangle$, see Sec.~\ref{sec3b}, 
 resulting in a robust and quantized $\pi$ shift.

\subsection{Charging effects}
\label{sec4c}

Finally we turn to an interacting version of our setup, where the floating Majorana island 
comes with a large charging energy, $E_C \gg \max \left( \varepsilon_C, \lambda_j \right)$.
We consider Coulomb valley conditions such that the physics is dominated by virtual transitions connecting the lowest island charge state ($Q$) to neighboring states with charge $Q \pm 1$ only.
A Schrieffer-Wolff transformation yields the effective cotunneling Hamiltonian
\begin{equation}\label{CB:H}
H(t) = \sum_{j} \varepsilon_j(t) \left( d_j^\dagger d_j^{} - \frac12 \right) + \sum_{j \neq k = 1}^4 \Lambda_{jk} d_j^\dagger d_k^{} \gamma_k \gamma_j,
\end{equation}
with  the cotunneling amplitudes 
$\Lambda_{j k} = 2 \lambda_j \lambda_k/ E_C$. Note that $H(t)$
conserves the total particle number of the dot subsystem. 

Non-adiabatic transitions
are then governed by an effective two-level Hamiltonian describing the respective level crossing.
With the transition amplitudes $\left( u_{jk}, v_{jk, n_j} \right)$, where $u_{jk} = u_{jk}^\ast$ and $v_{jk,1} = -v^\ast_{jk, 0} \equiv v_{jk}$,
we find that ${\cal S}(t_{A,B,C})$ acts on $|\bm n\rangle \otimes |\Phi\rangle$   as ($\bar n_j=1-n_j$)
\begin{eqnarray}
{\cal S}(t_A) &=& \delta_{n_1, n_3}  +
\delta_{n_1, \bar n_3} \Biggl[ u_{13} +v_{13, n_1}\\ \nonumber&\times&  \left(d_1^\dagger-d_1^{}\right)\gamma_1 \left(d_3^\dagger-d_3^{}\right) \gamma_3 \Biggr],
\end{eqnarray} 
and similarly for ${\cal S}(t_B)$ with $(1,3)\rightarrow (2,4)$.
Transitions near $t_{C}=t_D$ simplify since CAR processes are strongly suppressed under Coulomb blockade conditions.
We obtain  
\begin{widetext}
\begin{eqnarray} \nonumber
{\cal S}(t_C)&=& \delta_{n_1, n_4} \delta_{n_2 n_3}  +\delta_{n_1, \bar n_4} \delta_{n_2, n_3} \left [ u_{14} +
v_{14, n_1} \left(d_1^\dagger - d_1^{}\right)\gamma_1 \left(d_4^\dagger - d_4^{}\right)\gamma_4 \right] \\
&+& \delta_{n_1, n_4} \delta_{n_2, \bar n_3} \left[ u_{23}  +
v_{23, n_2}\left(d_2^\dagger - d_2^{}\right)\gamma_2 \left(d_3^\dagger - d_3^{}\right)\gamma_3  \right]  \\
&+& \delta_{n_1, \bar n_4} \delta_{n_2 ,\bar n_3} \left[ u_{14} +
v_{14, n_1}\left(d_1^\dagger - d_1^{}\right)\gamma_1 \left(d_4^\dagger - d_4^{}\right)\gamma_4  \right] \nonumber \times
\left[ u_{23} + v_{23, n_2}
\left(d_2^\dagger - d_2^{}\right)\gamma_2 \left(d_3^\dagger - d_3^{}\right)\gamma_3  \right] .
\end{eqnarray}
\end{widetext}
Since we have only ``uncorrelated'' transitions near $t_C$,
we here were able to go beyond perturbation theory in $\lambda_j$.

\begin{figure}\includegraphics[width=\columnwidth]{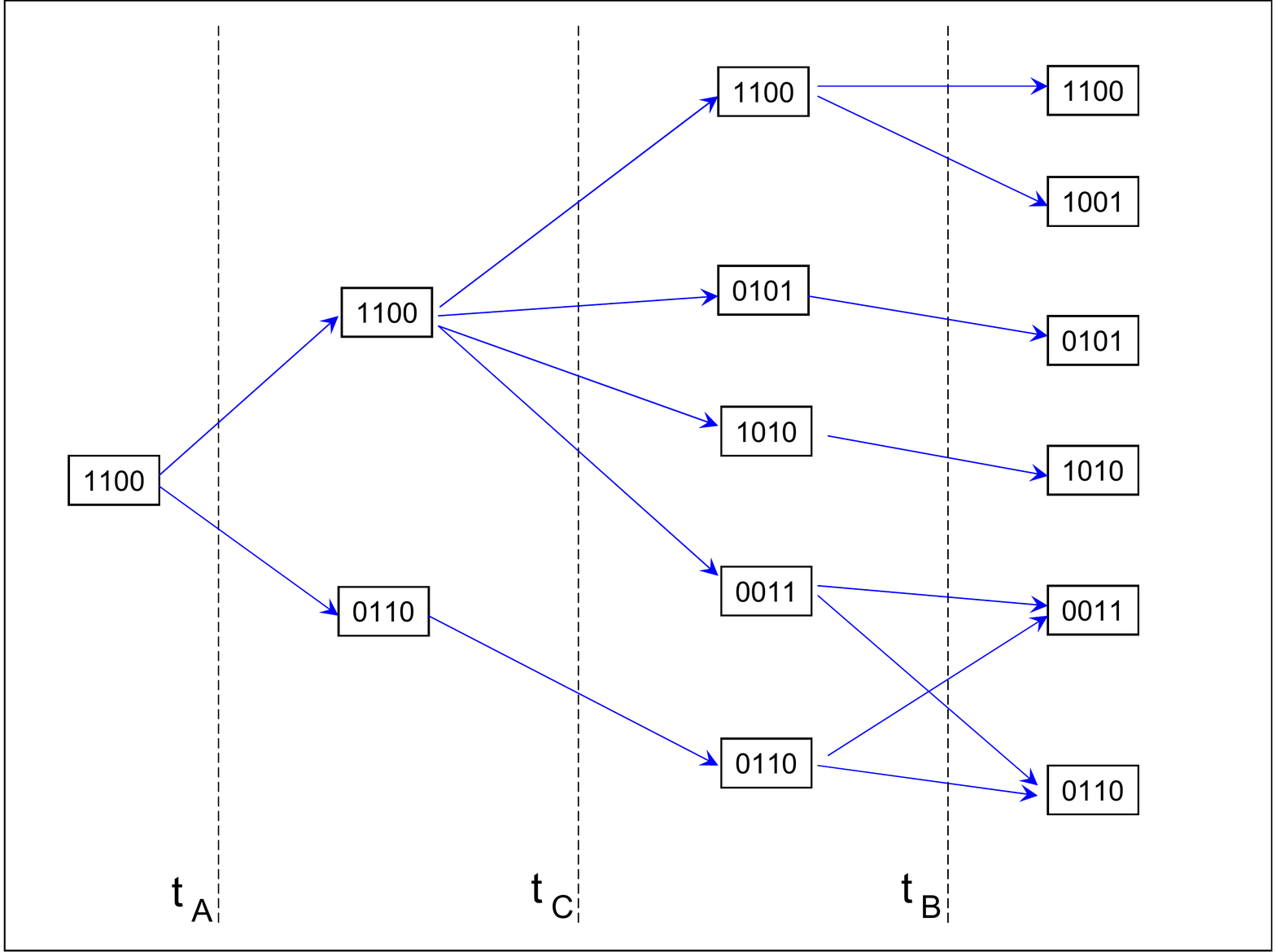}
\caption{\label{fig6}
Same as Fig.~\ref{fig5} but for a Coulomb-blockaded Majorana island, again for initial states with $\bm n^i=(1100)$. Note that in this case,
only particle-number conserving transitions are allowed.  
}
\end{figure}

Using the above rules, the probability distribution $P[\bm n^f]$
can be readily computed. Since
 only particle-number conserving transitions are allowed,
the diagrams in Fig.~\ref{fig6} represent a truncated version of the $E_C=0$ case discussed in 
Secs.~\ref{sec2} and \ref{sec3}.
For ${\bm n}^i = (1100)$, we find that only the following two probabilities show an interference pattern:
\begin{eqnarray}
{\cal P}_{0011} &=& \left| u_{13} v_{14} v_{23} u_{24} + e^{2 i \chi} v_{13} v_{24} \right|^2,\\  \nonumber
{\cal P}_{0110} &=& \left| u_{13} v_{14} v_{23} v_{24}^\ast + e^{2 i \chi} v_{13} u_{24} \right|^2.
\end{eqnarray}  
In particular, no $\pi$ shift is possible due to the absence of CAR processes.

\section{Conclusions and Outlook}\label{sec5}

 We have introduced a general scheme to address localized topological quasiparticles through controlled and robust features of many-body interferometry in the time-energy domain. 
 We believe that this approach offers many interesting perspectives for future research.  First, variants of the above setup may be implemented for studying more exotic quasiparticles. Our interferometric approach may thus open a new experimental window for probing topological excitations.  Second, when using many islands and a large number $N$ of dots with tunable level energies, 
 an extended lattice structure in time-energy space can be generated. 
By switching selected tunnel couplings to a small value at defined time intervals, one can control the links forming this lattice.  
One may then study percolation and phase transitions in such a lattice, similar to but different from recent works on random unitary circuits and quantum graphs \cite{Li2018,Nahum2018,Khemani2018}. 
Moreover, following earlier work on particle localization in energy-time space \cite{Landauer1987,Gefen1987,Lubin1990,Goldhirsch1991}, it is intriguing to generalize this physics to a many-body setting involving topological quasi-particles.  
Finally, with additional couplings between dots, it should be possible to implement Majorana braiding protocols \cite{Alicea2012} in the time-energy domain.

\begin{acknowledgements}
 We thank Kyrylo Snizhko for discussions.
This project has been funded by the Deutsche Forschungsgemeinschaft (DFG,
German Research Foundation), Projektnummer 277101999, TRR 183 (project
C01), by the DFG  under Germany's Excellence Strategy - Cluster of Excellence Matter
and Light for Quantum Computing (ML4Q) EXC 2004/1 - 390534769, 
by the Israel Science Foundation, and by the Minerva foundation.
\end{acknowledgements}


\begin{thebibliography}{10}
\bibitem{foot1}
The spin representation is based on the unity relation
$\gamma{}_j \gamma_j^\dagger = \gamma_j^\dagger \gamma^{}_j = 1$. This relation also holds for
parafermions with $\mathbb{Z}_k$ symmetry, where the ``spin operators'' corresponding to Eq.~\eqref{spin12} 
do not commute with each other, $\sigma_{j,+} \sigma_{l,\nu=\pm} = -e^{2i\pi \nu/k} \sigma_{l,\nu} \sigma_{j,+}$ for $j<l$, and 
a  rich interference pattern can be expected.
Finally, we note that the above unity relation is  not satisfied for Bogoliubov quasiparticles.
\end{thebibliography}
\end{document}